\documentclass[12pt]{iopart}
\usepackage{setstack}
\usepackage{iopams}
\usepackage{graphicx}

\begin{document}
\title[]{Parametric coupling between macroscopic quantum resonators}

\author{$^{1}$L. Tian, $^{2}$M. S. Allman, $^{2}$R. W. Simmonds}

\address{$^1$University of California, Merced, School of Natural Sciences, Merced, CA 95344, USA}
\address{$^2$National Institute of Standards and Technology, 325 Broadway St, Boulder,
CO 80305, USA}

\date{\today{}}

\begin{abstract}
Time-dependent linear coupling between macroscopic quantum
resonator modes generates both a parametric amplification also
known as a {}``squeezing operation" and a beam splitter operation,
analogous to quantum optical systems. These operations, when
applied properly, can robustly generate entanglement and squeezing
for the quantum resonator modes. Here, we present such coupling
schemes between a nanomechanical resonator and  a superconducting
electrical resonator using applied microwave voltages as well as
between two superconducting lumped-element electrical resonators
using a \mbox{r.f. SQUID}-mediated tunable coupler. By calculating
the logarithmic negativity of the partially transposed density
matrix, we quantitatively study the entanglement generated at
finite temperatures. We also show that characterization of the
nanomechanical resonator state after the quantum operations can be
achieved by detecting the electrical resonator only. Thus, one of
the electrical resonator modes can act as a probe to measure the
entanglement of the coupled systems and the degree of squeezing
for the other resonator mode.
\end{abstract}
\submitto{New Journal of Physics}
\maketitle
\normalsize

\section{Introduction}
The quantum behavior of macroscopic resonators has only recently
been demonstrated in solid-state systems and is currently under
intensive exploration. Electrical resonators such as a
superconducting transmission line resonator coupled to a single
Cooper pair box or phase qubits and a d.c. SQUID resonator
interacting with a flux qubit have shown features of single
microwave quanta \cite{Yale,NIST,UCSB,NTT}. Nanomechanical
resonators capacitively coupled to a single electron transistor
have been measured, approaching the quantum limit with femtometer
displacements
\cite{EResonator,KarlsruheRMP,MResonator,BlencowePR}. Recently, a
transmission line resonator coupled to a nanomechanical resonator
has been used to cool the nanomechanical motion to hundreds of
quanta \cite{NatPhysRegal}. Besides being a wonderful testing
ground for exploring quantum physics at the macroscopic level,
these systems can provide high-Q harmonic-oscillator networks for
quantum engineering and quantum information processing
\cite{BraunsteinRMP}.

Micro-fabricated resonators interact by electromagnetic forces. By
controlling the circuit parameters, various time-dependent
couplings can be generated to manipulate the quantum state of the
coupled system. In a previous work \cite{TianPRB2005}, one of us
(L.T.) studied the effective amplification of the coupling
amplitude between a nanomechanical resonator and a qubit by
parametrically pumping the qubit with a fast pulse. This scheme
can be applied to produce entanglement between resonator modes as
well as to generate Schr\"{o}dinger cat states in the
nanomechanical system. In a recent work \cite{JacobsPRL2007}, it
was shown that by controlling the coupling between a
nanomechanical resonator and an ancilla qubit, it is possible to
engineer an arbitrary Hamiltonian for the nanomechanical resonator
mode. In another work \cite{YuanPRA2007}, a pulse technique is
used to form arbitrary quantum states of a nanomechanical
resonator in the coupled resonator-qubit system. In addition,
experimental realization of parametric coupling has been achieved
for two coupled flux qubits \cite{NiskanenSCI2007}.

In this paper, we show that quantum features such as entanglement
and squeezing can be generated and tested using the Gaussian
states of macroscopic resonator modes coupled in a tunable way.
Parametric modulation of the coupling strength results in a
squeezing operation or a beam splitter operation, necessary
ingredients for engineering quantum features. Starting from a
separable initial state, inseparability (or entanglement) can be
generated by linear operations. Two physical systems are studied:
in one system a nanomechanical resonator is capacitively coupled
to a superconducting electrical resonator mode, in another system
two lumped-element electrical resonators are coupled to each other
through a tunable mutual inductance. In the first system, if we
start with both the nanomechanical and electrical resonator modes
in their ground states, careful control over squeezing and beam
splitter operations can produce squeezed states in the
nanomechanical mode. Note that previous work on creating squeezed
states \cite{YurkeSqueezing} in nanomechanical-resonator systems
uses quantum reservoir engineering, feedback control techniques,
and an (effective) nonlinear coupling
\cite{ZollerSqueezing,KorotkovSqueezing,MoonSqueezing,ZhouSqueezing}.
One advantage of the scheme described here is that linear coupling
between the solid-state resonators, realizable with current
technology, can be easily controlled in both magnitude and
frequency. In addition, our approach not only provides a method
for generating entanglement and squeezed states, but it also
provides a practical way for detecting the squeezing. We show that
for Gaussian states, complete information about the coupled
resonators can be obtained by only measuring the quadrature
variances of the electrical resonator mode \cite{MilburnSET} which
couples more strongly to the detector than the mechanical mode. In
this way, we can avoid any difficulties associated with directly
measuring the nanomechanical mode with high resolution. Here, the
electrical resonator plays the role of a knob that controls the
behavior of the nanomechanical resonator while also acting as a
detector to probe this behavior. Using a \mbox{r.f.
SQUID}-mediated tunable coupler, this work can be extended in a
straightforward way to a system with two coupled electrical
resonator modes. For typical dilution refrigerator temperatures
($\sim30$ mK), we can pick both resonator frequencies ($\sim10$
GHz) to operate above the quantum limit where the two electrical
resonators will definitely be cooled to their ground state, making
it easy to prepare these modes in squeezed states by linear
operations. Also, the two modes can be measured simultaneously,
which provides a direct observation of the cross correlations
between the two modes for a clear comparison with theoretical
predictions.

Entanglement is the key component for many quantum information
protocols such as quantum teleportation. Previous work
\cite{TianPRB2006} studied the quantum teleportation of
nanomechanical modes in a purely solid-state network, where high
fidelity could be achieved for the final state with the assistance
of a highly entangled two-mode squeezed vacuum state even at
finite resonator temperatures. The schemes studied in this paper
provide a detailed account of how to engineer and evaluate this
entanglement to ensure the success of the quantum teleportation
protocol, as well as to generate one important quantum state -- a
squeezed state of the resonators.

\section{Coupled macroscopic resonators}
In this section, we investigate the realization of linear, tunable
coupling between macroscopic resonator modes and the parametric
modulation of such coupling using a circuit architecture.

\subsection{a nanomechanical mode coupled to transmission line resonator}
First, consider a nanomechanical resonator capacitively coupled to
a resonant superconducting transmission line. Quantum behavior of
high-Q electrical modes has been demonstrated using
superconducting transmission lines \cite{EResonator}. The
eigenmodes of a one dimensional transmission line resonator of
length $L$ between $(-L/2,L/2)$ can be obtained by considering the
charge distribution along the transmission line,
\[
\theta(x,t)=\int_{-L/2}^{x}dx^{\prime}q(x^{\prime},t)\] where
$q(x^{\prime},t)$ is the linear charge density at the location
$x$. The lowest even mode has the voltage distribution
\begin{equation}
V(x)=\frac{1}{c}\frac{\partial\theta}{\partial
x}=\sqrt{\frac{\hbar\omega_{b}}{cL}}\cos\frac{2\pi
x}{L}\left(\hat{b}+\hat{b}^{\dagger}\right)
\label{eq:Vx}
\end{equation}
after the quantization of the variables. Here, the frequency of
this mode is $\omega_{b}=2\pi/L\sqrt{lc}$, with $l$ the inductance
per unit length and $c$ the capacitance per unit length of the
transmission line.

\begin{figure}
\center{
\includegraphics[%
  bb=54bp 72bp 570bp 690bp,
  clip,
  width=8cm]{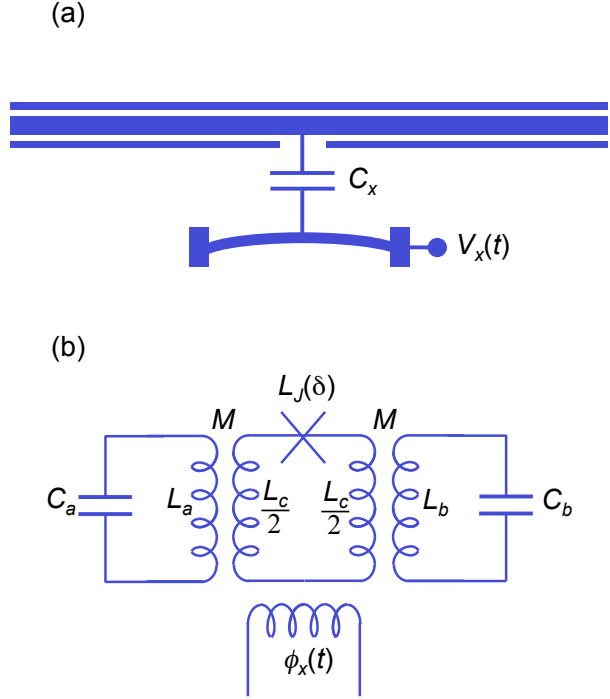}}
  
\caption{\label{figure1}(a) A nanomechanical resonator interacts
via a capacitance $C_{x}$ with a transmission line resonator. (b)
Two lumped-element electrical resonators interact via a \mbox{r.f
SQUID}-mediated tunable coupler.}
\end{figure}

The nanomechanical resonator is located at the middle of the
superconducting transmission line near $x=0$ which is a voltage
antinote of this even mode (figure~\ref{figure1}). The
nanomechanical resonator is coupled capacitively via a
displacement dependent capacitance $C_{x}$. To lowest order,
$C_{x}=C_{x}^{0}(1+\hat{x}_a/d_{0})$ depends linearly on the
displacement coordinate of the nanomechanical motion. The coupled
interaction can be derived as
\begin{equation}
H_{int}=\frac{1}{2}C_{x}^{0}(1+\frac{\hat{x}}{d_{0}})(V_{x}-V(0))^{2}
\label{eq:Hint}
\end{equation}
where the nanomechanical resonator is biased at $V_{x}$ and $d_0$
is an effective distance between the two resonator electrodes. We
study the behavior of the lowest nanomechanical mode with a
displacement $\hat{x}=\delta x_{0}(\hat{a}+\hat{a}^{\dagger})$,
where $\omega_{a}$ is the frequency of the mechanical mode, $m$ is
the effective mass and $\delta x_{0}=\sqrt{\hbar/2m\omega_{a}}$.
The interaction can be derived as
$H_{int}=-\frac{\hat{x}}{d_{0}}C_{x}^{0}V_{x}V(0)$, assuming that
the size of the nanomechanical resonator is much smaller than the
length of the transmission line. If we apply a voltage
$V_{x}(t)=2V_{x}^{0}\sin\omega_{d}t$ with a drive frequency
$\omega_{d}$, the linear interaction can be written as
$H_{int}=-2\lambda_{0}\sin\omega_{d}t(\hat{a}+\hat{a}^{\dagger})(\hat{b}+\hat{b}^{\dagger})$
with a coupling strength
\begin{equation}
\lambda_{0}=C_{x}^{0}V_{x}^{0}\frac{\delta
x_{0}}{d_{0}}\sqrt{\frac{\hbar\omega_{b}}{cL}}
\label{eq:lambda}
\end{equation}
By adjusting the driving frequency of the voltage, it is possible
to generate various linear operations, as discussed in more detail
in section \ref{sec3}. Hence, the modulation of the coupling
strength provides an effective tool for controlling the
entanglement of the coupled resonators.

\subsection{two coupled lumped-element resonator modes}
\label{2CoupRes} Alternatively, two electrical resonators can be
coupled together using a \mbox{r.f. SQUID}-mediated tunable
coupler. The circuit schematic (figure~\ref{figure1}) shows a
\mbox{r.f. SQUID} placed between two lumped-element
$LC$-resonators. The \mbox{r.f. SQUID} acts like an inductive
transformer where the Josephson junction provides a tunable
inductance $L_J(\delta) = \Phi_0/(2\pi I_o cos(\delta))$ and
$\delta$ is the phase difference across the junction, $I_o$ is the
junction critical current, and $\Phi_0$ is a flux quantum. The use
of small Josephson junctions with high current density allows us
to ignore the self capacitance $C_J$ of the junction so that we
remain in an operation regime where $\omega^2 L_J C_J\ll 1$ for
all the relevant frequencies $\omega$. Furthermore, by careful
design and operation, we can avoid any direct coupling between the
two resonators so that a single effective mutual inductance,
\begin{equation}
M_{eff}(\delta)=\frac{M^2}{L_c+L_J(\delta)}
\label{eq:Meff}
\end{equation}
derives only from the mutual inductive coupling $M$ between each
resonator and the central \mbox{r.f. SQUID} transformer coil with
geometrical self inductance $L_c$. This leads to a coupled
interaction of strength \cite{NJOPv7p230y2005},
\begin{equation}
H_{int}(\delta)=-M_{eff}(\delta)\sqrt{\frac{\hbar\omega_a}{2L_a}\frac{\hbar\omega_b}{2L_b}}(\hat{a}+\hat{a}^{\dagger})(\hat{b}+\hat{b}^{\dagger}),
\label{eq:Hint2}
\end{equation}
where $\omega_a$ and $\omega_b$ are frequencies of the two resonators respectively.

\begin{figure}\center{
\includegraphics[%
  bb=54bp 480bp 570bp 660bp,
  clip,
  width=12cm]{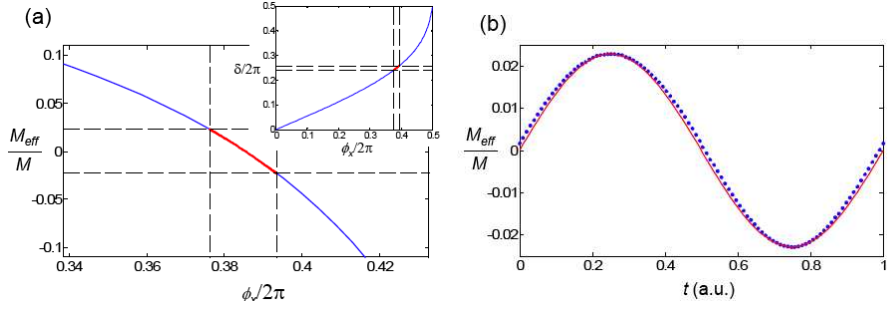}}

\caption{\label{figure2} (a) $M_{eff}/M$ versus external
$\phi_x/2\pi$, where the (red online) highlight is the region
where the coupling is modulated. The inset shows the dependence of
$\delta$ on $\phi_x$ for $\beta = 0.85$ (b) $M_{eff}/M$ (points)
versus time. The thin (red online) line is a sine function.}
\end{figure}

In order to modulate the coupling between the resonators, we must
modulate the phase difference $\delta$. This can be achieved using
a \mbox{r.f. bias} coil which applies an external flux $\phi_x =
2\pi\Phi_x/\Phi_o$ to the \mbox{r.f. SQUID} coil, thereby inducing
a current through the Josephson junction and thus changing the
phase difference $\delta$. The phase difference $\delta$ is
related to the external flux $\phi_x$ through flux quantization
such that $\beta \sin(\delta) + \delta - \phi_x = 0$, where $\beta
= L_c/L_J(0)$. This places some constraints on the design and
subsequent operation of the tunable coupler
\cite{NJOPv7p230y2005}. Namely, we would like to operate in a
regime where $\beta\lesssim1$ so that the relationship between
$\delta$ and $\phi_x$ remains non-hysteretic and the inductive
coupling passes from anti-ferromagnetic (AF) through zero to
ferromagnetic (FM). If we choose a particular \mbox{d.c. flux}
offset and a relatively small amplitude for the \mbox{r.f. flux}
modulation, we can achieve a roughly linear relationship between
the drive flux $\phi_x$ and the effective mutual inductance
$M_{eff}$ without any residual direct coupling. An example is
shown (figure~\ref{figure2}) for $\beta = 0.85$ and other device
parameters specified later. Thus, for a sinusoidal flux drive
$\phi_{x}(t)=\phi_{dc}-\phi_{rf}\sin(\omega_{d}t)$ at frequency
$\omega_d$, we can produce a sufficient parametric modulation of
the coupling strength (\ref{eq:Hint2}) and hence perform the
squeezing or beam splitter operations. With $\phi_{dc} = 2.4178$
and $\phi_{rf} = 0.0540$, we can closely approximate a sinusoidal
variation in the size of the effective mutual inductance
(figure~\ref{figure2}). Slight deviations from the ideal behavior
will only appear as negligibly small residual direct coupling or
small amplitude components at higher harmonic frequencies which do
not corrupt the squeezing or beam splitter operations.

A coupling magnitude similar to (\ref{eq:lambda}) can be derived
for this circuit with
\begin{equation}
 \lambda_0=\frac{\Delta M_{eff}}{2}\sqrt{\frac{\hbar\omega_a}{2L_a}\frac{\hbar\omega_b}{2L_b}}
 \label{eq:elambda}
\end{equation}
where $\Delta M_{eff}$ is the amplitude of the modulating
effective mutual inductance.

Coupling electrical resonators has several advantages: 1) the
coupling between the electrical resonators can be made stronger
fairly easily through simple design modifications, more so than
for coupling between a mechanical resonator and an electrical
resonator, 2) the requirements on device temperature for clearly
operating in the quantum regime have also been achieved
\cite{Yale,NIST,UCSB}, 3) electrical resonators have already
demonstrated sufficiently (high Q's) long coherence times
\cite{Yale,NIST,UCSB}, making is possible to study their dynamic
behavior, and 4) direct measurement of both resonators provides
more information on the overall quantum behavior of the coupled
system. Thus, even if pursuing the nanomechanical approach proves
to be difficult, it will be very promising to pursue coupled
electrical resonators in order to investigate the generation of
entanglement and other quantum physics.

\section{Linear operations by parametric coupling}
\label{sec3}
\subsection{coupling in the rotating frame}
Consider the time-dependent linear coupling between two resonator
modes of the form
 $H_{int}=-2\lambda_{0}\sin\omega_{d}t(\hat{a}+\hat{a}^{\dagger})(\hat{b}+\hat{b}^{\dagger})$,
where $\lambda_0$ is the coupling amplitude and $\omega_d$ is the
modulation frequency of the coupling strength. Here the operators
$\hat{a}$ ($\hat{a}^\dagger$) and  $\hat{b}$ ($\hat{b}^\dagger$)
are the annihilation (creation) operators for the resonant mode.
At the driving frequency $\omega_{d}=\omega_{b}-\omega_{a}$, the
coupling in the interaction picture has the form
\begin{equation}
H_{I}=i\lambda_{0}(\hat{a}^{\dagger}\hat{b}-\hat{a}\hat{b}^{\dagger})\label{eq:Hbs}
\end{equation}
which  generates a beam splitter operation in a similar to fashion
to that found in quantum optics. At the driving frequency
$\omega_{d}=\omega_{b}+\omega_{a}$, the coupling has the form
\begin{equation}
H_{I}=i\lambda_{0}(\hat{a}\hat{b}-\hat{a}^{\dagger}\hat{b}^{\dagger})
\label{eq:sq}
\end{equation}
which generates a squeezing operation and hence entanglement
between the two modes. These two operations are sufficient for
manipulating entanglement and for generating squeezed states of
the resonators \cite{EisertPRL2004} as we will discuss in detail
below.

We focus on the Gaussian states of the coupled resonator modes. A
Gaussian state can be fully characterized by the covariance
matrix:
\begin{equation}
(\sigma)_{ij}=\frac{1}{2}\langle\hat{x}_{i}\hat{x}_{j}+\hat{x}_{j}\hat{x}_{i}\rangle-[\hat{x}_{i},\hat{x}_{j}]\label{eq:sig_def}
\end{equation}
with $i,j=1,2,3,4$. The operators in this expression are the
quadrature variables of the resonators with
$\hat{x}_{1}=\hat{x}_{a},\hat{x}_{2}=\hat{p}_{a},\hat{x}_{3}=\hat{x}_{b},\hat{x}_{4}=\hat{p}_{b}$
for the modes $a$ and $b$ respectively. Here $\langle\rangle$ is
the ensemble average for the density matrix of the Gaussian states
and $[,]$ is the commutator between the variables.  Note that
linear operations and dissipation due to white noise maps Gaussian
states to Gaussian states \cite{GaussianState}.

The generic form of the covariance matrix for coupled resonators is
\begin{equation}
\sigma  =\left(\begin{array}{cc}
A & C\\
C^{T} & B\end{array}\right)
\label{eq:sig}
\end{equation}
where $A$ ($B,C$) is $2\times2$ diagonal matrix with the elements
$a_{1}$ and $a_{2}$,
\begin{equation}
A =\left(\begin{array}{cc}
a_{1} & 0\\
0 & a_{2}\end{array}\right)
\label{eq:ABC}
\end{equation}
(and similarly for $b_{i}$ and $c_{i}$). We set the linear
displacement to be $\langle\hat{x}_{i}\rangle=0$, which does not
affect the entanglement of the system. To describe the finite
temperature of the resonators, we use the effective temperature
index $\Theta_{\alpha}=\coth\frac{\hbar\omega_{\alpha}}{k_{B}T}$
for $\alpha=a,b$. When $k_{B}T\ll\hbar\omega_{\alpha}$, we have
$\Theta_{\alpha}=1$ and the mode is in its ground state. The
initial covariance matrix of the thermal states for the uncoupled
resonators is a diagonal matrix $\sigma_{0}$ with the diagonal
elements $(\Theta_{a},\Theta_{a},\Theta_{b},\Theta_{b})/4$ where
the quadrature variances of the resonators are $\langle
x_{i}^{2}\rangle=\langle p_{i}^{2}\rangle=\Theta_{i}/4$.

\subsection{linear operations: squeezing and beam splitter}
It can be shown that the squeezing operation in (\ref{eq:sq})
performs the following transformation on the covariance matrix,
\begin{equation}
\sigma=e^{A_{sq}r_{2}}\sigma_{0}e^{A_{sq}r_{2}}
\end{equation}
where\[
A_{sq}=\left(\begin{array}{cccc}
0 & 0 & 1 & 0\\
0 & 0 & 0 & -1\\
1 & 0 & 0 & 0\\
0 & -1 & 0 & 0\end{array}\right)\] is the squeezing operator and
$r_{2}=\lambda_{0}t$ is the squeezing parameter over a duration
$t$. We find that  \[ e^{A_{sq}r_{2}}=\left(\begin{array}{cccc}
\cosh r_{2} & 0 & \sinh r_{2} & 0\\
0 & \cosh r_{2} & 0 & -\sinh r_{2}\\
\sinh r_{2} & 0 & \cosh r_{2} & 0\\
0 & -\sinh r_{2} & 0 & \cosh r_{2}\end{array}\right).\]

Another linear operation, the beam splitter type of operation, on
the resonator modes in (\ref{eq:sq}), performs the following
transformation on the covariance matrix,
\begin{equation}
\sigma(\varphi)=e^{A_{bm}^{T}\varphi}\sigma_0 e^{A_{bm}\varphi}
\label{eq:rot}
\end{equation}
where $\varphi=\lambda_0 t$ and \[
A_{bm}=\left(\begin{array}{cccc}
0 & 0 & 1 & 0\\
0 & 0 & 0 & 1\\
-1 & 0 & 0 & 0\\
0 & -1 & 0 & 0\end{array}\right).\]
We can show that\[
e^{A_{bm}\varphi}=\left(\begin{array}{cccc}
\cos\varphi & 0 & -\sin\varphi & 0\\
0 & \cos\varphi & 0 & -\sin\varphi\\
\sin\varphi & 0 & \cos\varphi & 0\\
0 & \sin\varphi & 0 & \cos\varphi\end{array}\right).\] This
transformation produces a beam splitter operation on the coupled
resonators. In particular, it swaps the states of the two modes
when $\phi=\pi/2$.

\section{Entanglement and squeezing of the coupled resonators}
In this section, we study the quantum engineering of entanglement
between coupled macroscopic resonator modes and the squeezing of
the resonators using the parametric coupling circuits discussed in
the previous sections.

\subsection{entanglement}
At zero temperature, the squeezing operation in (\ref{eq:sq})
generates squeezed vacuum states between two resonator modes which
are also entangled states \cite{Separability}. Here, we show that
even at finite temperatures, entanglement can be generated by the
squeezing operation starting from an initial state with the
covariance matrix $\sigma_0$. After a duration $t_{sq}$ with the
coupling magnitude $\lambda_0$, the squeezing operation transforms
the elements of the covariance matrix to
\begin{eqnarray}
a_{1}=a_{2} & = & \frac{1}{4}\left(\Theta_{a}\cosh^{2}\lambda_{0}t_{sq}+\Theta_{b}\sinh^{2}\lambda_{0}t_{sq}\right)\nonumber \\
b_{1}=b_{2} & = & \frac{1}{4}\left(\Theta_{a}\sinh^{2}\lambda_{0}t_{sq}+\Theta_{b}\cosh^{2}\lambda_{0}t_{sq}\right)\label{eq:ele-sq}\\
c_{1}=-c_{2} & = &
\frac{1}{4}\left(\Theta_{a}+\Theta_{b}\right)\sinh\lambda_{0}t_{sq}\cosh\lambda_{0}t_{sq}\nonumber
\end{eqnarray}which increase with the squeezing parameter
$r_{2}=\lambda_{0}t_{sq}$. Note that the above relation shows
$c_{i}\leq\sqrt{a_{i}b_{i}}$ for $i=1,2$ with the equality valid
for a large squeezing parameter $r_2$.

\begin{figure}\center{
\includegraphics[%
  bb=54bp 330bp 540bp 540bp,
  clip,
  width=12cm]{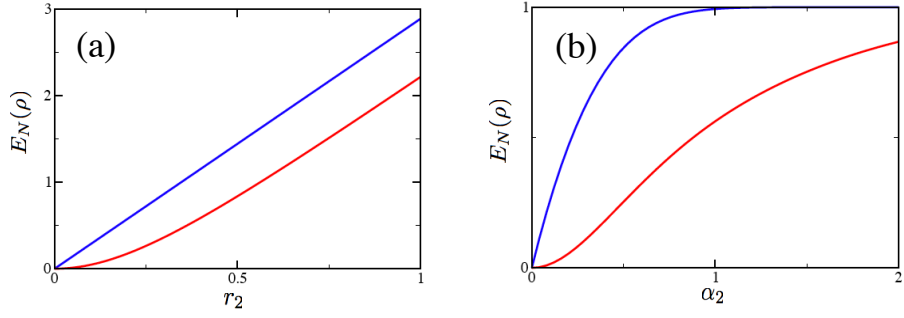}}

\caption{\label{figure3} (a) The logarithmic negativity
$E_{N}(\rho)$ for the two mode squeezed states versus the
squeezing parameter. (b) The logarithmic negativity $E_{N}(\rho)$
for effective polarization entanglement state $\rho_{pl}$ in
(\ref{eq:rhopl}) versus displacement parameter $\alpha_2$  with
$\alpha_1=3$. In both plots, the upper (blue online) curve is for
zero temperature with $\Theta_{a}=\Theta_{b}=1$ and the lower (red
online) curve is for finite temperature with $\Theta_{a}=4$ and
$\Theta_{b}=1$. }
\end{figure}

To quantitatively characterize the entanglement for the above
mixed state, we calculate the logarithmic negativity
\cite{VidalNegativity}, $E_{N}(\rho)=\log_{2}||\rho^{T_{A}}||$, of
the partially transposed density matrix of this two-mode
continuous variable Gaussian state. For Gaussian states, the
logarithmic negativity can be derived directly from the
covariances:
\begin{equation}E_{N}(\rho)=\sum_{\alpha=1}^{2}F(c_{\alpha})
\label{En}
\end{equation}
where the function $F(c_\alpha)$ is\[
F(c_\alpha)=\{\begin{array}{cc}
0\, & \rm{for}\:2c_\alpha\geq1\\
-\log_{2}(2c_\alpha)\, & \rm{for}\:2c_\alpha<1\end{array}\] and
the variables $c_{\alpha}$ can be derived from the roots of the
following equation:\[
x^{4}+4(\det(A)+\det(B)-2\det(C))x^{2}+16\det(\sigma)=0.\] The
matrices $A,B,C$ were defined previously \ref{eq:ABC} for the
covariance matrix $\sigma$ \ref{eq:sig}.

In figure~\ref{figure3} (a), we plot $E_N(\rho)$ versus the
squeezing parameter $r_2$. At zero temperature ($\Theta_{a,b}=1$),
the entanglement increases nearly linearly with $r_2$. At finite
temperature, finite squeezing is required to overcome the thermal
fluctuations before any entanglement can be generated. With large
squeezing parameter $r_2 > 0.5$, the logarithmic negativity can be
approximated as
\begin{equation}
E_{N}(\rho)\approx\frac{1}{\log2}\left(2\lambda_{0}t_{sq}-\log\frac{2\Theta_{a}\Theta_{b}}{\Theta_{a}+\Theta_{b}}\right),
\label{eq:En-rsq}
\end{equation}
increasing linearly with the squeezing parameter. At $r_2=0.5$,
$\Theta_a=4$ for the nanomechanical mode, and $\Theta_b=1$ for the
electrical mode, we have $E_N(\rho)=0.76$ giving finite
entanglement. Hence, \emph{at finite temperature}, entanglement
between coupled modes can be generated by applying large
squeezing.

Here, we want to compare the two types of entanglement as apposed
to the generation of entanglement between resonator modes as
described in Ref. \cite{TianPRB2005} using a different method. For
the case of a resonator coupled to a solid-state qubit, parametric
pulses can be used to flip the qubit state every half period of
the resonator mode, $\pi/\omega_{a,b}$, producing an effective
amplification of the resonator displacements\cite{NISTunp,UCSB}.
For two resonators in a pure state, the entangled state can be
expressed as $|\psi\rangle \propto
|\alpha_1,\alpha_2\rangle+|-\alpha_1,-\alpha_2\rangle$. This state
is comparable to the entangled state between two qubits
$|0,0\rangle +|1,1\rangle$ and we call it the effective
polarization entangled state. For the mixed state at finite
temperature, the density matrix of the entangled state can be
written as
\begin{equation}
\rho_{pl}\propto (D_{12}+D_{12}^\dagger) \rho_0 (D_{12}+D_{12}^\dagger)
\label{eq:rhopl}
\end{equation}
where $D_{12}$ is the displacement operator
\[
D_{12}=e^{\alpha_2^\star \hat{a}-\alpha_2 \hat{a}^\dagger}
e^{\alpha_1^\star \hat{b}-\alpha_1\hat{b}^\dagger}
\]
and we neglect the normalization factor in $\rho_{pl}$. The
logarithmic negatively for this state (figure~\ref{figure3}) has
also calculated. After a rapid increase with $\alpha_2$, the
logarithmic negativity saturates at $E_N(\rho)\rightarrow 1$. At
finite temperature, we also have $E_N(\rho)\rightarrow 1$ when
$\alpha_2 \gg \sqrt{k_BT/\hbar\omega_a}$.

\subsection{squeezing}

\begin{figure}\center{
\includegraphics[
  bb=54bp 189bp 558bp 582bp,
  clip,
  width=10cm]{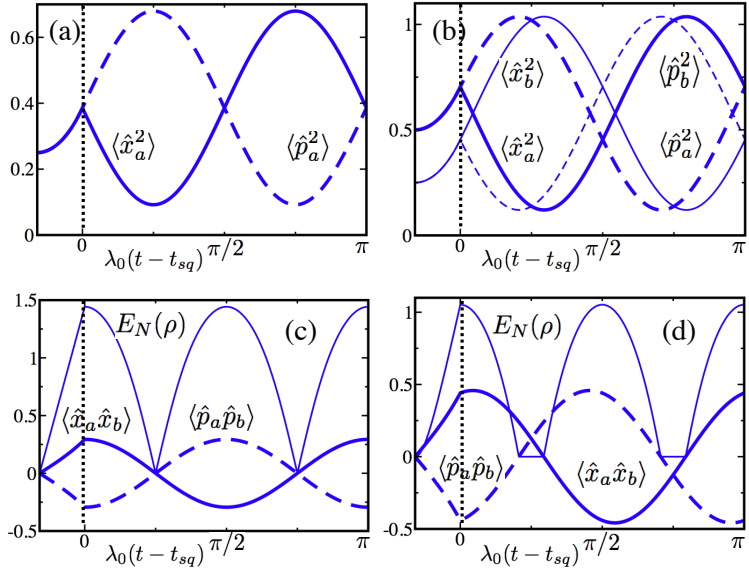}}

\caption{\label{figure4}The covariances and the logarithmic
negativity $E_{N}(\rho)$ for (a, c) $\Theta_{a}=\Theta_{b}=1$ and
(b, d)  $\Theta_{a}=2$ and $\Theta_{b}=1$. The dashed line at
$t=t_{sq}$ signifies the end of the \emph{squeezing} operation
with $t_{sq}=0.5/\lambda_0$. Following squeezing, the \emph{beam
splitter} operation is applied for $\pi/\lambda_0$.}
\end{figure}

Following the squeezing operation in (\ref{eq:ele-sq}),
entanglement between the resonators can be manipulated or adjusted
using \emph{a subsequent beam splitter operation}. Applying the
beam splitter operation for a duration $t-t_{sq}$, the covariances
can be expressed as,
\begin{eqnarray}
\widetilde{a}_{1} & = & \frac{a_{1}+b_{1}}{2}-\frac{b_{1}-a_{1}}{2}\cos2\varphi-c_{1}\sin2\varphi\nonumber \\
\widetilde{b}_{1} & = & \frac{a_{1}+b_{1}}{2}+\frac{b_{1}-a_{1}}{2}\cos2\varphi+c_{1}\sin2\varphi\label{eq:dyn}\\
\widetilde{c}_{1} & = &
c_{1}\cos2\varphi+\frac{a_{1}-b_{1}}{2}\sin2\varphi\nonumber
\end{eqnarray} with $\varphi=\lambda_{0}(t-t_{sq})$, and similar
relations can be derived for
$\widetilde{a}_{2},\widetilde{b}_{2},\widetilde{c}_{2}$. One
interesting feature is that the covariance matrix is divided into
a direct sum of two subsets between the variables
$\{\hat{x}_{a},\hat{x}_{b}\}$ and $\{\hat{p}_{a},\hat{p}_{b}\}$
respectively. Here, the covariances show oscillatory behavior
depending on the applied time $t-t_{sq}$ or the phase $\varphi$
\emph{of the applied beam splitter operation}, as is shown in
figure~\ref{figure4}.

Starting at maximum entanglement  at $t=t_{sq}$, we see
$E_{N}(\rho)$ decreases to zero near $t=t_{sq}+\pi/4\lambda_0$
($\varphi=\pi/4$) when the two resonator modes become separable,
then returns to its maximum value at $t=t_{sq}+\pi/2\lambda_0$
($\varphi=\pi/2$). At $t=t_{sq}+\pi/4\lambda_0$, we have
\begin{eqnarray}
\widetilde{a}_{1}=\widetilde{b}_{2} & = & (\Theta_{a}+\Theta_{b})e^{-2\lambda_{0}t_{sq}}/8\nonumber \\
\widetilde{a}_{2}=\widetilde{b}_{1} & = &
(\Theta_{a}+\Theta_{b})e^{2\lambda_{0}t_{sq}}/8\label{eq:var-sq}\end{eqnarray}
showing a significant amount of squeezing for the quadrature
fluctuations $\widetilde{a}_{1}=\langle \hat{x}_a^2\rangle $ and
$\widetilde{b}_{2}=\langle \hat{p}_b^2\rangle $. The cross
correlation at this point is
$\widetilde{c}_{1}=\widetilde{c}_{2}=(\Theta_{a}-\Theta_{b})/8$.
At finite temperature, as plotted in figure~\ref{figure4} (d), the
negativity decreases to zero and the coupled resonators are in a
separable state for a finite interval near
$t=t_{sq}+\pi/4\lambda_0$ \cite{Separability}. With increasing
temperature, the duration of this interval increases as well,
where the entanglement is diminished by thermal fluctuations.
Meanwhile, the correlation between the logarithmic negativity and
the covariance elements $c_{1,2}$ can also be seen in
figure~\ref{figure4} (c) and (d).

In the special situation of zero temperature with
$\Theta_{a}=\Theta_{b}$, as plotted in figure~\ref{figure4} (a)
and (c), \emph{squeezed states} can be generated for the resonator
modes. Here we have $\widetilde{a}_{1}=\widetilde{b}_{2}$,
$\widetilde{a}_{2}=\widetilde{b}_{1}$ and
$\widetilde{c}_{1}=-\widetilde{c}_{2}$ \emph{at all times}. When
$t=t_{sq}+\pi/4\lambda_0$, the cross correlations as well as the
entanglement vanish with $\widetilde{c}_{1,2}=E_{N}(\rho)=0$. When
both modes are initially prepared in their ground states with
$\Theta_{a,b}=1$, $\sqrt{\widetilde{a}_{1}\widetilde{a}_{2}}=1/4$.
Thus, a squeezed state is generated in \emph{each} resonator. For
the coupled system of a nanomechanical resonator and an electrical
resonator, this provides a novel way of generating squeezed states
in the nanomechanical mode using linear parametric coupling.

\section{Detection and the covariance matrix}
A crucial requirement for studying the quantum properties of
coupled resonators is the detection and verification of any
engineered entanglement. Below we show that by only measuring the
quadrature variances of one resonator -- the electrical resonator
-- the full covariance matrix can be constructed.

In reality, measurement of the vibration of a nanomechanical mode
is limited by the weak coupling between that mode and the
detector. The electrical mode, in contrast, contains an
electromagnetic signal that couples more strongly to the detector
than the nanomechanical mode. Hence, the electrical mode provides
an effective probe of the nanomechanical mode, which reduces the
demand on the detector efficiency. For coupled electrical
resonators, both modes can be measured simultaneously to directly
test the theoretical results presented above.

The variances of the $\{\hat{x}_{a},\hat{x}_{b}\}$ quadratures at
any time during the beam splitter operation are determined by
three initial parameters: $a_{1},\, b_{1},\, c_{1}$, and similarly
for the $\{\hat{p}_{a},\hat{p}_{b}\}$ quadratures. Measurements of
the variances $\widetilde{b}_{1}$ (and $\widetilde{b}_{2}$), at
three different times (in three sets of experiments with the beam
splitter operation applied for different $\varphi$) that are
linearly independent, can provide complete information about the
entanglement dynamics of the coupled modes. For example, we choose
to make measurements at $\varphi=0,\,\pi/4,\,\pi/2$ as defined
above. At $\varphi=0$, $\widetilde{b}_{1}=b_{1}$ is measured; at
$\varphi=\pi/2$, $\widetilde{b}_{1}=a_{1}$ is measured; at
$\varphi=\pi/4$, $\widetilde{b}_{1}=\frac{a_{1}+b_{1}}{2}+c_{1}$
is measured, from which $c_{1}$ can be obtained by combining the
previous results. Hence, the variances of the coupled modes can be
uniquely determined by the measurements of the quadratures of (one
of) the electrical modes. The times at which the measurements are
performed can be adjusted. By choosing
$\varphi=\pi/8,\,\pi/4,\,3\pi/8$, the quadrature fluctuations
exceed thermal noise with $\widetilde{b}_{1}\gg\Theta_{b}/4$ in
all three measurements as is shown in figure~\ref{figure4} (a, b).
Note that to measure the squeezed state of a nanomechanical mode,
a fast beam splitter operation with a phase $\pi/2$ can be applied
before measuring the electrical mode. This operation transfers the
states between the nanomechanical mode and the electrical mode, so
that the subsequent measurements of the electrical resonator
provide direct information of the nanomechanical mode.

This method provides a useful way of detecting the {}``hard''
mechanical mode by detecting the {}``easy'' electrical mode.
Mechanical modes are in general {}``hard'' to measure because they
couple very weakly to detectors. By transferring the state of the
mechanical mode to the electrical mode, which is in general
{}``easy'' to detect, we can clearly access the dynamical features
of the mechanical mode. This state transfer technique was
developed previously for phase qubits coupled to an electrical
resonator \cite{NIST}. Use of this method yields important
information about entanglement by taking advantage of a dynamical
process to reduce the requirements on the detector efficiency.

For the superconducting transmission line resonator, phase
sensitive detection of the quadrature variables has been performed
to reveal the vacuum Rabi splitting due to resonator-qubit
coupling by measuring the transmission or reflection of a resonant
signal \cite{EResonator} and for a resonator coupled to a
nanomechanical beam to study the fluctuation of the quadrature
variables approaching their quantum limit
\cite{RegalLehnert0801.1827,NatPhysRegal}. Alternatively, some
work has studied the phase sensitive detection of resonator modes
with a single electron transistor (SET) by  mixing the signal of
the resonator with a large radio-frequency reference signal and
coupling the mixed signal to a SET detector
\cite{MilburnSET,ClelandSET}. The nonlinear response of the SET
enables a high resolution measurement of the electrical resonator
\cite{SchoelkoptSET}. The sensitivity of the measurement, however,
can be limited by the detector noise. In the discussion above,
only the detection of the covariance matrix of the coupled
resonators was studied. Full characterization of the resonator
states can be obtained by a quantum state tomography method as has
been developed in the context of quantum optics
\cite{LeibfriedRMP2003}.

\section{Discussion and Conclusions}
The effects studied in this paper can be tested with realistic
parameters. Typical parameters for a nanomechancial resonator
\cite{MResonator} are $\omega_{a}/2\pi=100\,\rm{MHz}$,
$Q_{a}=10^{4}$, $d_{0}=50\,\rm{nm}$, and $\delta
x_{0}=5\,\rm{fm}$. For a superconducting transmission line
resonator \cite{EResonator}, $\omega_{b}/2\pi=5\,\rm{GHz}$,
$cL=4\,\rm{fF}$, and $Q_{b}=10^{4}$. With a bias voltage of
$V_{x}^{0}=4\,\rm{V}$ and a coupling capacitance of
$C_{x}^{0}=0.65\,\rm{fF}$, we find $\lambda_{0}/2\pi=6\,\rm{MHz}$
and the linear operations can be performed over a characteristic
time scale of $t_{sq}= \pi/4\lambda_{0}\sim130\,\rm{ns}$. For two
coupled electrical resonators, choosing reasonable parameters for
lumped-element quantum circuits \cite{JunctionQubit}
$\omega_{a}/2\pi=9\,\rm{GHz}$, $\omega_{b}/2\pi=10\,\rm{GHz}$,
$L_a=L_b=5L_c=10M=500\,\rm{pH}$, $I_o=2.8\,\mu\rm{A}$, and a
modulating flux drive like that described in section
\ref{2CoupRes} gives $\lambda_{0}/2\pi=6\,\rm{MHz}$ so that the
linear operations can also be performed for this system over a
timescale of $t_{sq}\sim130\,\rm{ns}$.

We didn't discuss the effect of decoherence on the dynamics of the
coupled resonator systems  \cite{decoherence}. Many discussions
can be found in the literature. The finite quality factors of both
the electrical resonator and the nanomechanical resonator can
limit the entanglement or squeezing in this scheme. Here, for a
squeezing parameter $r_2=0.5$, we have $\bar{n}\sim 2 \langle
\hat{x}_b^2\rangle \sim 1$. Assuming a modest quality factor of
$Q=10^{4}$ which is experimentally realizable, the decoherence
time is on the order of $2\pi Q/(\bar{n}\omega_{b})\sim 2\,\mu
\rm{s}\gg$~$t_{sq}$ sufficiently long enough to observe the
entanglement and squeezing generated in these systems.

For the measurement process, we proposed a scheme using homodyne
detection of the covariance matrix of the superconducting
electrical resonator mode by applying the beam splitter operation
at three different durations giving a full account of the
covariance matrix of the coupled resonators. This avoids the
difficulty of a direct measurement of the nanomechancial resonator
mode. Considering a coupling capacitance of $C_{x}^{o}\sim
0.5\,\rm{fF}$ and a maximum possible bias voltage of $V_{x} \sim
10\,\rm{V}$, the demands on the detector efficiency are great if
we consider directly measuring the quantum limited fluctuations of
the nanomechanical resonator. For two coupled electrical
resonators, direct detection on both resonators together can
reveal the covariances for all controlled operations, extremely
useful for testing these types of systems and the schemes we have
described for generating entanglement and squeezed states.

In conclusion, we studied the generation of controllable
entanglement and squeezing in coupled macroscopic resonators, in
particular, a nanomechanical-electrical resonator, by applying
parametrically modulated linear coupling. Two systems are studied
in detail: the coupling between a nanomechanical resonator and a
superconducting transmission line resonator, and the coupling
between two superconducting electrical resonators. The parametric
coupling is calculated in both cases. Furthermore, effective
detection of the entanglement and squeezing in one of the modes
can be achieved by measuring the other (electrical) mode. We have
considered specific, reasonable operating parameters and find that
both of these systems should be experimentally feasible. Squeezing
of the nanomechanical system, although difficult, should be
possible at very low temperatures whereas the relatively simple
two coupled electrical resonator system should show clear
squeezing for a variety of typical operating conditions. The
scheme studied here is a key component for entanglement based
quantum information protocols such as quantum teleportation in a
solid-state network \cite{TianPRB2006}.

\section*{Acknowledgement}
R.W. S. is
supported by NIST and IARPA under Grant No. W911NF-05-R-0009.

\section*{References}

\end{document}